\documentclass[authoryear,5p,times,round,colon]{elsarticle}
\usepackage{booktabs}
\usepackage{graphicx} 
\usepackage{amsmath,epsfig,amssymb,subfigure,bm,dsfont}
\usepackage{lipsum} 
\usepackage{array} 
\usepackage{multirow} 
\usepackage{multicol}
\usepackage{booktabs} 
\usepackage{longtable} 
\usepackage{tablefootnote}
\usepackage{algorithm}  
\usepackage{algpseudocode}  
\usepackage{amsmath}  
\usepackage{algorithmicx}
\usepackage{xcolor}
\usepackage[english]{babel}
\usepackage{multirow}
\usepackage{graphicx}
\usepackage{booktabs}
\usepackage{diagbox}
\usepackage{tabularx}
\usepackage{colortbl}
\usepackage{tikz}
\usepackage{tcolorbox}
\usepackage{booktabs}

\usepackage{tikz}

\journal{Journal of \LaTeX\ Templates}

\begin{document}

\begin{frontmatter}

\title{Optimizing Student Ability Assessment: A Hierarchy Constraint-Aware Cognitive Diagnosis Framework for Educational Contexts}

\author[1,3,4]{Xinjie Sun}
    \ead{xinjiesun@mail.ustc.edu.cn}
      \author[2,3]{Qi Liu \corref{mycorrespondingauthor}}
    \ead{qiliuql@ustc.edu.cn}
    \author[1,3]{Kai Zhang}
    \ead{kkzhang08@ustc.edu.cn}
    \author[5]{Shuanghong Shen}
    \ead{closer@mail.ustc.edu.cn}
    \author[6]{Fei Wang}
    \ead{wang-fei@nus.edu.sg}
    \author[1,3]{Yan Zhuang}
    \ead{zykb@mail.ustc.edu.cn}
    \author[1,3]{Zheng Zhang}
    \ead{zhangzheng@mail.ustc.edu.cn}
    \author[1,3,4]{Weiyin Gong}
    \ead{weiyingong@mail.ustc.edu.cn}
    \author[7,3]{Shijin Wang}
    \ead{sjwang3@iflytek.com}
    \author[4]{Lina Yang}
    \ead{linayang@lpssy.edu.cn}
    \author[4]{Xingying Huo}
    \ead{12120040@bjtu.edu.cn}

    \cortext[mycorrespondingauthor]{Corresponding authors: Qi Liu.}

    \address[1]{School of Computer Science and Technology, University of Science and Technology of China, Hefei, China}
   \address[2]{School of Artificial Intelligence and Data Science, University of Science and Technology of China, Hefei, China}
    \address[3]{State Key Laboratory of Cognitive Intelligence, Hefei, China}
    \address[4]{School of Computer Science, Liupanshui Normal University, Liupanshui, China}
    \address[5]{Institute of Artificial Intelligence, Hefei Comprehensive National Science Center, Hefei, China}
    \address[6]{School of Computing, National University of Singapore, Singapore, Singapore}
    \address[7]{iFLYTEK AI Research (Central China),  Wuhan, China}

\begin{abstract}

Cognitive diagnosis (CD) aims to reveal students' proficiency in specific knowledge concepts. With the increasing adoption of intelligent education applications, accurately assessing students' knowledge mastery has become an urgent challenge. Although existing cognitive diagnosis frameworks enhance diagnostic accuracy by analyzing students' explicit response records, they primarily focus on individual knowledge state, failing to adequately reflect the relative ability performance of students within hierarchies. To address this, we propose the \emph{\textbf{H}ierarchy Constraint-Aware \textbf{C}ognitive \textbf{D}iagnosis Framework (\textbf{HCD}),} designed to more accurately represent student ability performance within real educational contexts. Specifically, the framework introduces a hierarchy mapping layer to identify students’ levels. It then employs a hierarchy convolution-enhanced attention layer for in-depth analysis of knowledge concepts performance among students at the same level, uncovering nuanced differences. A hierarchy inter-sampling attention layer captures performance differences across hierarchies, offering a comprehensive understanding of the relationships among students' knowledge state. Finally, through personalized diagnostic enhancement, the framework integrates hierarchy constraint perception features with existing models, improving the representation of both individual and group characteristics. This approach enables precise inference of students' knowledge state. Research shows that this framework not only reasonably constrains changes in students' knowledge states to align with real educational settings, but also supports the scientific rigor and fairness of educational assessments, thereby advancing the field of cognitive diagnosis.

\end{abstract}


\end{frontmatter}


\section{INTRODUCTION}

\begin{figure}
    \centering
    \includegraphics[width=0.5\textwidth]{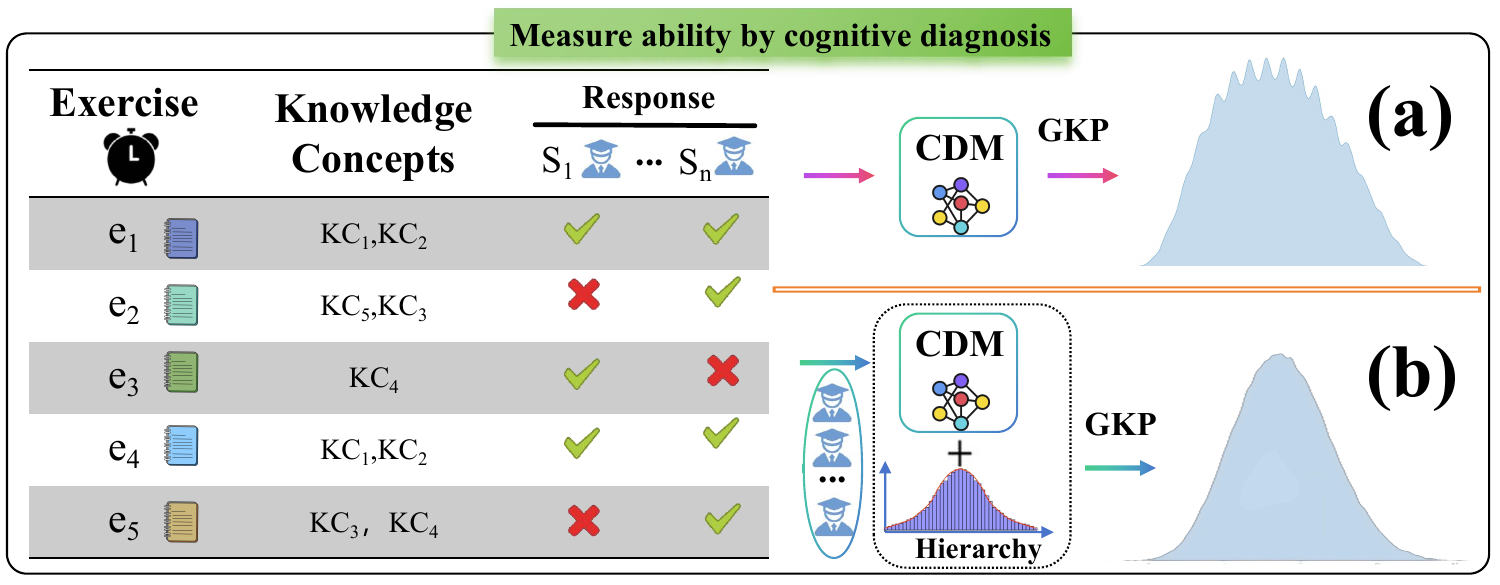}
    \caption{\textbf{A toy example of cognitive diagnosis, where CDM stands for cognitive diagnosis model and GKP stands for group knowledge proficiency.}}\label{figureIntro}
\end{figure}

Cognitive diagnosis (CD) is a core task in educational data mining, aimed at revealing students' proficiency in specific knowledge concepts (KCs) by analyzing their response logs \citep{bib022,bib009}. With the widespread adoption of intelligent education applications, the demand for cognitive diagnosis in assessing and improving individual development is increasingly growing \citep{bib046}. By analyzing students' interaction data with exercises, cognitive diagnosis models students' abilities and evaluates their knowledge state, thereby enhancing learning efficiency and helping students gain a clearer understanding of their learning progress. As shown in Figure \ref{figureIntro} (a), cognitive diagnosis follows common paradigms that are widely applied in various fields, including gaming \citep{bib051}, medical diagnosis \citep{bib052}, and education \citep{bib008,bib054}.

Cognitive diagnosis models (CDM) have undergone several stages of development, and their quality relies on the design of interactive functions that simulate the complex relationships between students and knowledge concepts. Although traditional Item Response Theory (IRT) \citep{bib055} and Multidimensional Item Response Theory (MIRT) \citep{bib021} provide a foundation for CDM, their assumptions limit practical applications. The DINA model \citep{bib038}, as an important form of CDM, offers in-depth analysis of students' knowledge mastery, aiding in the formulation of personalized learning strategies. In recent years, research that integrates deep neural networks \citep{bib047} and graph neural networks \citep{bib056,bib057} has significantly enhanced overall performance in various applications.



Despite the fact that existing cognitive diagnosis frameworks have improved diagnostic accuracy to some extent by fully utilizing students' explicit response records (e.g. exercise difficulty \citep{bib058}, related knowledge concepts \citep{bib059}, and exercise texts \citep{bib060}) and taking into account special cases during the learning process ((e.g. guessing behavior \citep{bib061}), they mainly focus on individual students' knowledge mastery levels. However, we believe that current research fails to adequately reflect students' relative hierarchical ability performance within groups. This limitation may lead to a lack of comprehensiveness in inferring ability levels across various knowledge concepts. Therefore, accurately assessing students' relative hierarchical performance within groups to ensure that assessments are reasonable and aligned with the demands of real educational contexts remains an urgent issue, as illustrated in Figure \ref{figureIntro} (b). Addressing this challenge will advance the field of cognitive diagnosis and highlight the scientific and equitable nature of contemporary educational assessments.


To more accurately diagnose each student's cognitive state, this paper explores the key challenges of incorporating hierarchy constraint perception features into cognitive diagnosis. \textbf{First}, effectively defining student hierarchies within real educational contexts is a fundamental issue. \textbf{Second}, although students at the same level may be similar in overall ability, their performance on specific knowledge concepts can differ significantly, necessitating a detailed analysis to accurately assess their knowledge mastery. \textbf{Third}, comprehensively capturing performance differences among student groups requires that ability estimates remain precise within a reasonable range. \textbf{Finally}, since students' ability performance is influenced by both group and individual characteristics, enhancing cognitive diagnosis through personalization and establishing intrinsic connections between individual and group features is also an important research direction.




To address the aforementioned challenges, we propose the {{H}ierarchy Constraint-Aware {C}ognitive {D}iagnosis Framework ({HCD}).} This framework encompasses several key innovations aimed at enhancing the effectiveness and interpretability of cognitive diagnosis. \textbf{First}, we introduce a Hierarchy Mapping Layer to identify students' levels within real educational contexts. \textbf{Second}, we utilize a Hierarchy Convolution-Enhanced Attention Layer to conduct in-depth analyses of the performances of students at the same level across various knowledge concepts, effectively uncovering differences within similarities. \textbf{Third}, a Hierarchy Inter-Sampling Attention Layer is employed to capture performance differences among student groups, allowing for a comprehensive understanding of the relationships between the knowledge state of students at different levels through sampling attention. \textbf{Finally}, by leveraging Personalized Diagnostic Enhancement, we integrate hierarchy constraint perception features with existing cognitive diagnosis frameworks to enhance the representational capabilities of both individual and group characteristics, enabling effective inferences about students' knowledge state.

In summary, we present solutions to the challenges outlined above. Through this innovative approach, we can more reasonably constrain the variations in students' knowledge state. The main contributions of this research are as follows:

\begin{itemize}
\item We propose the Hierarchy Constraint-aware for Cognitive Diagnosis framework\footnote{To support reproducible research, we have published the data and code at https://github.com/xinjiesun-ustc/HCCD, encouraging further innovation in this field.}, effectively limiting the variations in students' knowledge state to make them more reasonable and aligned with real educational contexts.
\item We introduce attention mechanisms for both intra- and inter-level knowledge states, significantly enhancing the diagnostic accuracy of hierarchy constraint perception.
\item We employ a joint training mechanism that enhances the application effectiveness of personalized diagnosis, improving the interpretability and inferential capabilities of cognitive diagnosis.
\end{itemize}

\section{RELATED WORKS}
Accurately identifying students' knowledge state in real educational contexts is a highly valuable task. Subsequently, we will introduce relevant research work from two perspectives: cognitive diagnosis and educational hierarchy.
\subsection{Cognitive Diagnosis}








Cognitive diagnosis plays a critical role in various fields, including education, gaming, and healthcare. Its core objective is to deeply reveal learners' latent traits and cognitive characteristics by analyzing their testing data, thereby providing a modeling foundation for the interaction between learners and item features \citep{bib035,bib036}.

As one of the foundations of cognitive diagnosis, Item Response Theory (IRT) was initially proposed to describe the relationship between learners' performance on specific tasks and their latent abilities \cite{bib020,bib023}. By mapping item characteristics (e.g. difficulty and discrimination) to learner traits (e.g. ability), IRT models offer an effective statistical method for analyzing individual differences \citep{bib024,bib025}. However, IRT typically assumes that learners' cognitive traits are unidimensional, which limits its application in handling multidimensional cognitive tasks. \textbf{Our HCD framework integrates with IRT to form the HCD-IRT model.}

To address this limitation, Multidimensional Item Response Theory (MIRT) was introduced \cite{bib021}. MIRT models learners' cognitive traits as multidimensional vectors, providing a more accurate reflection of complex cognitive features. This innovative approach is widely used in educational assessment, enabling a more precise evaluation of learners' mastery across multiple knowledge concepts \citep{bib026,bib027,bib028}. \textbf{Our HCD framework integrates with MIRT to form the HCD-MIRT model.}

The DINA model has broad applications in educational assessment and personalized learning \citep{bib037}. The fundamental assumption of this model is that learners must possess mastery of all relevant knowledge concepts to successfully complete an item, and their responses are influenced by noise \citep{bib038}. It provides teachers with in-depth insights into students' knowledge mastery, aiding in the formulation of targeted teaching strategies \citep{bib039}. Additionally, researchers have extended the DINA model, incorporating multidimensional factors \citep{bib040} or integrating deep learning methods \citep{bib041}, to enhance the model's flexibility and applicability. \textbf{Our HCD framework integrates with DINA to form the HCD-DINA model.}

In recent years, researchers have begun exploring new avenues for cognitive diagnosis using advanced machine learning techniques \citep{bib030}. In these explorations, collaborative filtering and matrix factorization techniques have been employed to build learner ability models, improving the accuracy of exam performance predictions \citep{bib031,bib032,bib033}. Additionally, some studies \cite{bib029} have proposed neural network-based ability classifiers, significantly enhancing classification effectiveness and model stability by training on data from pre-trained attribute hierarchy models.

With advancements in deep learning technology, researchers are integrating these methods into cognitive diagnosis models to automatically learn the complex interactions between items and learner characteristics. NeuralCDM \citep{bib034} and DIRT \citep{bib022} are two representative studies in this field. NeuralCDM leverages the nonlinear modeling capabilities of neural networks to capture the complex relationships between learners and items in high-dimensional feature spaces, while DIRT combines the advantages of deep learning and traditional IRT models, enhancing predictive capability while retaining IRT's interpretability. Although these advances have improved the accuracy and adaptability of cognitive diagnosis, challenges remain in effectively modeling students' potential mastery of unseen items. \textbf{Our HCD framework integrates with NeuralCDM to form the HCD-NCDM model.}

Overall, the field of cognitive diagnosis is experiencing rapid innovation and development, with the combination of traditional theories and modern machine learning techniques offering new possibilities for the future of educational assessment.

\subsection{Educational Hierarchy}




In the field of education, research related to hierarchy involves multiple key theories and models \citep{bib001,bib002,bib003}. The hierarchy of needs theory proposed by \cite{bib004} emphasizes the impact of different levels of needs on learning motivation; cognitive development theory explores children's cognitive abilities at various developmental stages \citep{bib005}; and the taxonomy of educational objectives proposed by \cite{bib006} categorizes learning goals into levels such as knowledge, understanding, and application, providing guidance for curriculum design and assessment. Hierarchy analysis is used in educational evaluation and resource allocation, assisting in trade-offs among multi-level factors \citep{bib007}; meanwhile, hierarchy knowledge tracing models analyze students' mastery of knowledge, supporting personalized learning \citep{bib008,bib009}. Additionally, research on multi-level educational systems investigates the interactions within educational policies and practices, highlighting the importance of systems thinking, while studies on hierarchical learning environments focus on how different learning contexts affect student outcomes \citep{bib010,bib011}. These studies provide an important theoretical foundation and practical guidance for understanding hierarchy in education.

Effective stratification of educational objectives is crucial in the education sector. It not only helps educators and learners better understand learning goals but also enables more effective design and implementation of teaching strategies. Various methods exist to achieve this goal. For instance, Maslow's hierarchy of needs classifies human needs into five levels \citep{bib018}, while Gardner's theory of multiple intelligences emphasizes the importance of recognizing and developing students' diverse intelligences \citep{bib019}. \textbf{Our HCD framework is based on students' relative positions of average scores in prior statistics to conduct an initial hierarchy categorization.}

The application of educational hierarchy is wide-ranging, covering multiple areas. \cite{bib014} explores how different levels of learning objectives affect educational practices in their revised Bloom's taxonomy. \cite{bib015} developed a machine learning method to automatically label the matching of open educational resources and skill classifications, thereby addressing the labor-intensive issue of manual tagging and publicly released a pre-trained model, enhancing the practicality and policy impact of educational resources. Additionally, \cite{bib016} effectively improved high school students' computational thinking skills, learning motivation, and self-efficacy by stratifying teaching objectives through the integration of artificial intelligence education and the STEAM model. \cite{bib017} demonstrates how to effectively apply the principles of stratification in the design of innovative learning experiences. These studies collectively reveal the important role of educational stratification in enhancing teaching effectiveness and learning outcomes.

\section{PRELIMINARY}
In this section, we formally define the concept of Hierarchy Constraint-Aware Cognitive Diagnosis. Additionally, we introduce the key components of the HCD framework. Table 1 summarizes all the mathematical symbols used in this paper.
\begin{table}[t]
\centering
\caption{\textbf{Mathematical symbol and descriptions.}}
\small
\begin{minipage}{0.5\textwidth}
\begin{tabularx}{\textwidth}{l>{\raggedright\arraybackslash}X}
\toprule
Symbol & Description \\
\midrule
S, E, H & The set of students, exercises and hierarchies\\
KC, R   & The set of knowledge concepts and responses\\
Q-matrix &  Expert-annotated relationship matrix between exercises and knowledge concepts \\
$S_p^\theta$, $S_h^\theta$ & Personalized knowledge proficiency and knowledge proficiency under hierarchy\\
n, m, k, g   & The number of students, exercises, KCs, and hierarchies. \\
$\mathbf{H}_{intra},\mathbf{H}_{inter}$ & Feature vectors within and across hierarchies \\
\textbf{CEA, RSA } & Intra-Hierarchy Convolution-Enhanced Attention and Inter-Hierarchy Random Sampling Attention Layer \\
\bottomrule
\end{tabularx}
\end{minipage}
\end{table}

\subsection{Problem Definition}



In the framework of hierarchy constraint-aware cognitive diagnosis, we define several core sets: the  set of students \( S \), the  set of exercises \( E \), the  set of knowledge concepts \( KC \), the  set of hierarchy levels \( H \), and the  set of responses \( R \). Specifically, the students set \( S = \{s_1, s_2, \dots, s_n\} \) contains \( n \) distinct students, while the exercises set \( E = \{e_1, e_2, \dots, e_m\} \) includes \( m \) unique exercises. The knowledge concepts set \( KC = \{kc_1, kc_2, \dots, kc_k\} \) comprises \( k \) different concepts. The hierarchy levels set \( H = \{h_1, h_2, \dots, h_g\} \) consists of \( g \) levels. The responses set \( R = \{0, 1\} \) is used to evaluate students' answers, where \( 1 \) denotes a completely correct response and \( 0 \) denotes an incorrect one.

We construct the student test record \( L \) as a set of triples \( (s, e, h) \). Where \( s \) represents the student (\( s \in S \)), \( e \) denotes the exercise (\( e \in E \)), and \( h \) indicates the level the student is in (\( h \in H \)). Additionally, the exercise \( Q \)-matrix we use is predetermined (typically annotated by experts) and is denoted as \( Q = Q_{ij}^{m \times k} \). In this matrix, if exercise \( e_i \) is associated with knowledge concept \( kc_j \), then \( Q_{ij} = 1 \); otherwise, \( Q_{ij} = 0 \). Based on these definitions, we provide the specific definition of hierarchy-aware cognitive diagnosis as follows:

\textbf{Hierarchy Constraint-Aware Cognitive Diagnosis (HCD)}: Given a student's test records \( L = \{(s_n, e_1, h_g), (s_n, e_2, h_g), \ldots, \\(s_n, e_t, h_g)\} \) and the known \( Q \)-matrix, the goal is to infer the student's ability levels across various knowledge concepts. By introducing a hierarchy structure, we constrain the ability estimation process to ensure the assessment results remain reasonable and align with real educational contexts.

\begin{figure*}
    \centering
    \includegraphics[width=0.85\textwidth]{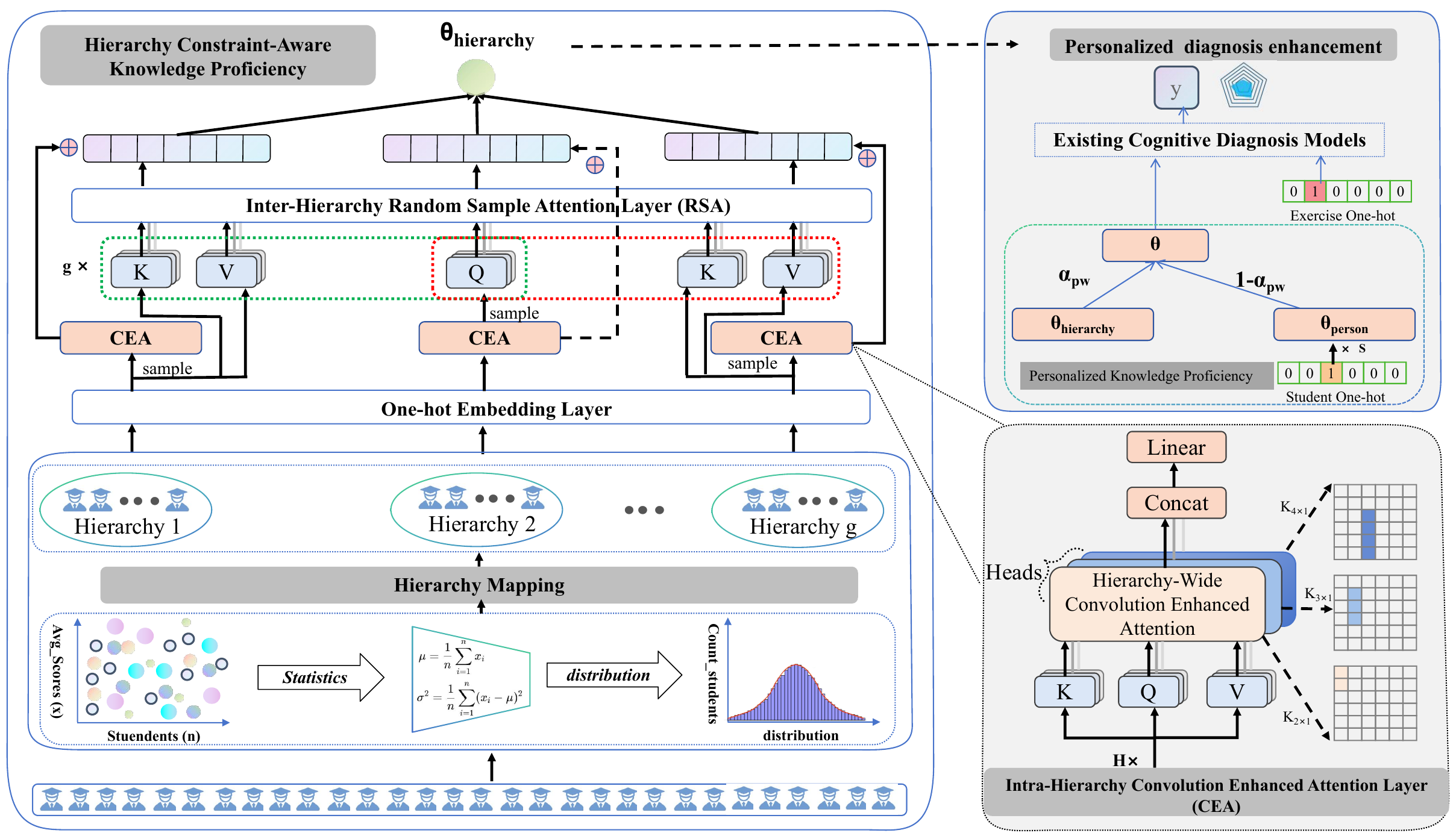}
    \caption{\textbf{Overview of HCD Framework.}}\label{figure02}
\end{figure*}
\subsection{Basic Factors}


Before starting the interactive analysis of cognitive diagnosis, it's important to outline some basic factors to help understand the interactions involved.

Firstly, the personalized knowledge proficiency vector \( S_p^\theta \) and the hierarchy constrained knowledge proficiency \( S_h^\theta \) are both vectors of equal length to the number of knowledge concepts. These two vectors are calculated by multiplying the one-hot encoding of the student ID \(\mathbf{x}_s\) and the one-hot encoding of the hierarchy ID \(\mathbf{x}_h\) with the trainable knowledge proficiency matrices \(\mathbf{S}\) and \(\mathbf{H}\):


\begin{equation}
    \begin{aligned}
    &S_p^\theta = \sigma(\mathbf{x}_s \times \mathbf{S}),\\
    &S_h^\theta = \sigma(\mathbf{x}_h \times \mathbf{H}),
    \end{aligned}
\end{equation}
where \( S_p^\theta \in (0, 1)^{1 \times k} \) represents the personalized knowledge proficiency of the student, and \( S_h^\theta \in (0, 1)^{1 \times k} \) indicates the proficiency of knowledge concepts under hierarchy awareness. The one-hot encoding of the student is denoted as \(\mathbf{x}_s \in \{0, 1\}^{1 \times n}\), with the corresponding trainable proficiency matrix given by \( \mathbf{S} \in \mathbb{R}^{n \times k} \). Similarly, the one-hot encoding for different hierarchies is represented as \(\mathbf{x}_h \in \{0, 1\}^{1 \times g}\), and its trainable proficiency matrix is \( \mathbf{H} \in \mathbb{R}^{g \times k} \). Where \(\sigma(\cdot)\) denotes the sigmoid activation function.

Secondly, when addressing the fundamental factors related to exercises, we focus on the difficulty of the knowledge concepts and the discrimination of the questions by one-hot encoding the exercise ID \(\mathbf{x}_e\). Within the framework of cognitive level research, the Q-matrix serves as one of the key elements to describe the relationship between exercises and knowledge concepts \citep{bib042}. The knowledge concepts involved in each exercise are typically quantified using this matrix. For a specific exercise \( e \), we can extract its corresponding knowledge concept relevance vector \(\mathbf{Q}_e\) from the Q-matrix, calculated as follows:

\begin{equation}
    \begin{aligned}
    &\mathbf{Q}_e = \mathbf{x}_e \times Q,
    \end{aligned}
\end{equation}
where \(\mathbf{x}_e \in \{0, 1\}^{1 \times m}\) represents the one-hot encoding of the ID of exercise \( e \), where \(\mathbf{Q}_e \in \{0, 1\}^{1 \times k}\) indicates the knowledge concept relevance vector for that exercise, effectively capturing the association between each exercise and its related knowledge concepts.

\section{ HCD Framework}

In this section, we will provide a comprehensive overview of the designed two-stage Hierarchy Constraint-Aware Cognitive Diagnosis (HCD) framework based on the formulaic expressions of the cognitive function. Subsequently, we will delve into the key details and implementation mechanisms of the two-stage model within this framework.

\subsection{Model Overview}



Typically, a student's performance on an exercise can be represented by the formula:

\begin{equation}
    \begin{aligned}
    & r = \text{CDM}(S^\theta, \psi_e),
    \end{aligned}
\end{equation}
where \( r \) denotes the student's response (e.g. score or correctness), \( S^\theta \) represents the student's knowledge mastery level, and \( \psi_e \) consists of parameters related to the exercise (e.g., difficulty \(\mathbf{h_e}^{\text{diff}} \in (0, 1)^{1 \times k}\) and discrimination \(h_e^{\text{disc}} \in (0, 1)\), with \(k\) being the number of knowledge concepts). This process is typically based on manually designed cognitive behavior models, such as the item response functions used in IRT.

To ensure that the predicted student ability values more accurately reflect the students' true capabilities, we consider relative constraints by comparing their performance with that of other students. For example, a student's ability value should not be excessively elevated due to relatively good performance in a single exam without considering their overall performance, especially when they may be underperforming in other areas. We divide knowledge mastery \( \theta \) into two components:

\begin{equation}
    \begin{aligned}
    & \theta = \mathcal{F}(S_p^\theta, S_h^\theta),
    \end{aligned}
\end{equation}
where \( S_h^\theta \) represents the relative ability characteristics under hierarchy constraints, while \( S_p^\theta \) denotes the student's personalized absolute performance characteristics. The function \( \mathcal{F} \) reflects the influence of these two characteristics.

As shown in Figure \ref{figure02}, the overall structure of the HCD framework can be divided into two main stages: the Hierarchy Constraint-Aware Modeling Stage and the Personalized Diagnostic Enhancement Stage. In the Hierarchy Constraint-Aware Modeling Stage, we design two attention networks to simulate the impact of hierarchy constraints on knowledge mastery. In the Personalized Diagnostic Enhancement Stage, we integrate personalized knowledge mastery (absoluteness) and knowledge mastery under hierarchy constraints (relativeness) to output predicted scores using the cognitive behavior function. After training with students' logs, we obtain each student's knowledge mastery as the diagnostic result. The following sections will provide a detailed introduction to these two stages.

\subsection{Hierarchy Constraint-Aware Modeling}

In the Hierarchy Constraint-Aware Modeling Stage, we designed two hierarchy attention networks to simulate the relationships in knowledge mastery among students both within and across hierarchies. The entire network consists of three components: hierarchy mapping, intra-hierarchy attention layer, and inter-hierarchy attention layer.

\subsubsection{Hierarchy Mapping}


Based on the hierarchy information derived from score intervals, valuable prior knowledge can be provided for the assessment process, allowing for more accurate capture of characteristics among students at different levels. This hierarchy information not only enhances the stability and rationality of ability assessments but also strengthens the overall robustness of evaluations, enabling better handling of students at varying levels and avoiding situations of “not enough” or “falling behind.” To more reasonably assess students' ability levels, we construct prior hierarchy information based on score intervals.

Let \( \overline{\mathcal{X}} = \{ \overline{x_1}, \overline{x_2}, \ldots, \overline{x_n} \} \) represent the average scores of all students, where \(\overline{x_i}\) denotes the average score of the \(i\)-th student. The mean \(\overline{x}\) and standard deviation \(\sigma\) of the average scores are calculated as follows:
\begin{equation}
    \begin{aligned}
    &\overline{x} = \frac{1}{n} \sum_{i=1}^{n} \overline{x_i}, \\
    &\sigma = \sqrt{\frac{1}{n} \sum_{i=1}^{n} (\overline{x_i} - \overline{x})^2}.
    \end{aligned}
\end{equation}

Based on this, we define the score intervals \(\mathcal{B} = \{ b_0, b_1, \dots, b_g \}\) and their corresponding label set \(\mathcal{L} = \{ l_0, l_1, \dots, l_g \}\). The mapping function \( f: \overline{\mathcal{X}} \to \mathcal{L} \) is defined as:
\begin{equation}
\begin{aligned}
& f(~\overline{x_i}~) = l_j \Leftrightarrow   b_j \leq \overline{x_i} < b_{j+1}, \, \forall i \in [1, n], \, j \in [0, g].
\end{aligned}
\label{eq:hierarchy}
\end{equation}

\subsubsection{One-hot Embedding}


This layer assigns trainable embeddings to each hierarchy entry \( h_j \), mapping them to potential knowledge proficiency features based on the student's current level. Let \( O \in \mathbb{R}^{n \times g} \) be the One-Hot encoding matrix, defined as:

\begin{equation}
\begin{aligned}
& O_{ij} = \begin{cases} 
1 & \text{if } \text{Eq.(\ref{eq:hierarchy})} \\ 
0 & \text{otherwise} 
\end{cases}, \quad \forall i \in [1, n], \, j \in [0, g-1],
\end{aligned}
\end{equation}
Subsequently, we convert the One-Hot encoding \( O \) into a knowledge concept vector \({H_{KCs}} \in \mathbb{R}^{n \times k} \) through a linear transformation, represented as:

\begin{equation}
\begin{aligned}
&{H_{KCs}} = W^o \cdot O  + b,
\end{aligned}
\end{equation}
where \( W^o \in \mathbb{R}^{g \times k} \) is the weight matrix of the linear layer, \( b \in \mathbb{R}^{k} \) is the bias vector, \( g \) is the number of hierarchy levels, and \( k \) is the number of knowledge concepts.

\subsubsection{Intra-Hierarchy Convolution Enhanced Attention Layer (CEA)
}

In educational contexts, students at the same level often exhibit significant overall ability similarities. However, despite being classified into the same level, their performances on specific knowledge concepts can vary considerably \citep{bib044}. A detailed analysis of the performances of students within the same level on identical knowledge concepts is crucial for accurately assessing their mastery of these concepts \citep{bib045}. Therefore, focusing on the performances of students at the same level on the same knowledge concepts enhances the model's sensitivity to knowledge mastery and improves the accuracy and effectiveness of educational cognition.

To better capture the performances of students at the same level on identical knowledge concepts, we first extract feature representations \(\mathbf{C}_h \in \mathbb{R}^{G' \times k}\) using different convolution kernels \( \mathbf{K}_h \):

\begin{equation}
\begin{aligned}
& \mathbf{C}_h = \mathcal{C}({H_{KCs}}; \mathbf{K}_h),
\end{aligned}
\end{equation}
where \( \mathcal{C} \) denotes the convolution operation, \( {H_{KCs}} \) is the knowledge concept embedding matrix, and \( G' \) is the dimension of the features after convolution.

Subsequently, we calculate the representations of queries \(\mathbf{Q}_h  \in \mathbb{R}^{G' \times k}\), keys \( \mathbf{K}_h \in \mathbb{R}^{G' \times k} \), and values \(\mathbf{V}_h \in \mathbb{R}^{G' \times k}\) for each convolution head \( h \), as follows:

\begin{equation}
\begin{aligned}
& \mathbf{Q}_h &= \mathbf{C}_h {W}_Q^h , \\
& \mathbf{K}_h &= \mathbf{C}_h {W}_K^h , \\
& \mathbf{V}_h &= \mathbf{C}_h {W}_V^h ,
\end{aligned}
\end{equation}
where \( {W}_Q^h, {W}_K^h, {W}_V^h \in \mathbb{R}^{k \times k} \) are the trainable weight matrices for queries, keys, and values.

To generate the attention output \( \mathbf{Z}_h \), we perform similarity measurements based on the features generated by convolution:

\begin{equation}
\begin{aligned}
\mathbf{Z}_h = \sigma\left(\frac{\mathbf{Q}_h \mathbf{K}_h^T}{\sqrt{k}}\right) \mathbf{V}_h,
\end{aligned}
\end{equation}
where \( \sigma(\cdot) \) is the Sigmoid function, emphasizing the interdependence among them.

Finally, we aggregate the attention outputs from all convolution heads to obtain the knowledge feature representation of students within the same level. This aggregation further enhances the expressive capability of the feature representations and helps the model better capture the complex relationships between different knowledge concepts. After aggregation, we need to apply further transformations to the intra-level feature representations to obtain \(\mathbf{H}_{\text{intra}} \in \mathbb{R}^{G \times k}\). The specific formulas are as follows:

\begin{equation}
\begin{aligned}
& \mathbf{H}_{\text{intra}}' = \frac{1}{H_A} \sum_{h=1}^{H} \mathbf{Z}_h ,\\
& \mathbf{H}_{\text{intra}} = {W}_{\text{intra}} \cdot \mathbf{H}_{\text{intra}}' + {b}_{\text{intra}} ,
\end{aligned}
\end{equation}
where \( H_A \) is the number of attention heads, \( {W}_{\text{intra}} \in \mathbb{R}^{k \times k} \) is the weight matrix of the fully connected layer, and \( {b}_{\text{intra}} \in \mathbb{R}^{k} \) is the bias term.

\subsubsection{Inter-Hierarchy Random Sample Attention Layer (RSA)}










The CEA module models the interactions among the same knowledge concepts within the same level, but capturing the performance differences among student groups to gain a comprehensive understanding of students' knowledge state remains a significant challenge. This is particularly true in sparse educational datasets, where maintaining reasonable estimates of students' abilities is crucial \citep{bib046}. By correlating the hierarchy distributions of different students, we can avoid extreme ability estimation issues.

In the inter-level attention mechanism, since students at the same level exhibit significant similarities, we randomly sample one student's features from other levels for interaction to reduce computational complexity. This design not only helps to decrease the computation load but also ensures the acquisition of information from students at different levels, enhancing the model's generalization ability. The specific definitions are as follows:

\begin{equation}
\begin{aligned}
& \mathbf{Q}_g &= {W}_Q \mathbf{H}_{\text{intra}}^g, \\
&\mathbf{K}_{g'} &= {W}_K \mathbf{H}_{\text{KCs}}^{g'}, \\
&\mathbf{V}_{g'} &= {W}_V \mathbf{H}_{\text{KCs}}^{g'},
\end{aligned}
\end{equation}
where \( g \) denotes the current level, and \( g' \) represents other levels.

Subsequently, to highlight the relevance of knowledge between levels and make the ability assessment of students more sensitive, we calculate the attention scores between the queries of level \( g \) and the keys of the sampled subset of students at level \( g' \) to obtain context-aware representations:

\begin{equation}
\begin{aligned}
\mathbf{Z}_{g}^{g'} = \text{softmax}\left(\frac{\mathbf{Q}_g \mathbf{K}_{g'}^{\top}}{\sqrt{d}}\right) \mathbf{V}_{g'}^{g'}.
\end{aligned}
\end{equation}
This process enhances the comprehensiveness and accuracy of student ability assessments by integrating information from different levels.

Following that, we average the sampled contextual representations from multiple other levels \( g' \) to consolidate the contextual information from all other levels:

\begin{equation}
\begin{aligned}
\mathbf{H}_{inter} = \frac{1}{|\mathcal{G}_{\text{others}}|} \sum_{g' \in \mathcal{G}_{\text{others}}} \mathbf{Z}_{g}^{g'},
\end{aligned}
\end{equation}
where \( |\mathcal{G}_{\text{others}}| \) represents the total number of other levels used for the averaging.

This design effectively integrates knowledge information from multiple levels, further enhancing the model's robustness and its ability to assess students' knowledge state.

Finally, we combine the outputs of intra-level attention with the sampled inter-level contextual information to obtain the final representation of hierarchy knowledge proficiency:

\begin{equation}
\begin{aligned}
\mathbf{ \theta_{hierarchy}} = \frac{1}{|\mathcal{H}|} \sum_{g' \in \mathcal{H}}(\mathbf{H}_{\text{intra}} + \mathbf{H}_{inter}),
\end{aligned}
\end{equation}
where \( |\mathcal{H}| \) is the number of hierarchies. This combination not only utilizes the contextual information within levels but also introduces inter-level interactions, ensuring that the model's estimates of student abilities are more comprehensive and accurate.

\subsection{Personalized Diagnosis Enhancement}


In this phase, we integrate hierarchical features with students' personalized knowledge levels. First, each student's personalized characteristics are described by a latent vector \( \theta_{\text{person}} \). Then, we adopt an adaptive optimization strategy for personalized weighting, combining hierarchical features with personalized features to form a comprehensive state for the student:

\begin{equation}
\begin{aligned}
& \alpha_{\text{pw}} = \sigma\left(\mathbf{E_{\text{person}}}[s]\right), \\
& \theta = \alpha_{\text{pw}} \cdot \theta_{\text{hierarchy}} + (1 - \alpha_{\text{pw}}) \cdot \theta_{\text{person}},
\end{aligned}
\end{equation}
where \( \mathbf{E_{\text{person}}} \) is a trainable weight matrix, and \( s \) is the student index.

In this model, we use the basic paradigm of cognitive diagnosis models to predict students' responses to exercises, with the specific formula given by: 
\begin{equation}
\begin{aligned}
&r = \mathbf{Q}_e \circ \left( \theta - \mathbf{h_e}^{\text{diff}} \right) \times h_e^{\text{disc}},
\end{aligned}
\end{equation}
this approach is applicable to various cognitive diagnosis models, such as IRT, MIRT, DINA and NeuralCD.

\begin{equation}
\begin{aligned}
&r = \text{CDM}(\theta, \psi_e),
\end{aligned}
\end{equation}
where \( \psi_e \) includes exercise-related parameters such as difficulty \( \mathbf{h_e}^{\text{diff}} \) and discrimination \( h_e^{\text{disc}} \).
\subsection{Objective Function}




To optimize all parameters within the  HCD  framework, we systematically analyze each response logs record in the test data and employ a cross-entropy loss function to measure the difference between the predicted values \( y \) and the students' actual response labels \( r \). The specific form of the loss function is as follows:

\small
\begin{equation}
    \begin{aligned}
   \mathcal{L} (y, r)  = -\sum_{i} \left( r_i \log(y_i) + (1 - r_i) \log(1 - y_i) \right).
    \end{aligned}
\end{equation}
\normalsize

We utilize the Adam optimization algorithm to minimize this loss function. By separately evaluating the outputs based on hierarchical features, personalized features, and existing prediction methods, we are able to capture students' learning state more comprehensively. This multi-level loss setup helps to more accurately identify students' ability differences across various knowledge domains, thereby optimizing the assessment effectiveness. The specific loss function definitions are as follows:

\begin{equation}
    \begin{aligned}
   & loss_{h} =  \mathcal{L} (y_{hierarchy}, r),\\
   & loss_{p} =  \mathcal{L} (y_{person}, r),\\
   & loss_{i} =  \mathcal{L} (y_{integration}, r),\\
   & loss =  loss_{h} +  loss_{p} +  loss_{i},
    \end{aligned}
\end{equation}
where \( y_{hierarchy} \) represents the prediction based on hierarchical features, \( y_{person} \) represents the prediction based on students' personalized learning characteristics, and \( y_{integration} \) denotes the output using existing prediction methods. The experimental section will further detail other setup specifics.

\begin{table}[h]
    \centering
    \caption{\textbf{Statistics of all datasets.}}
    \footnotesize  
    \resizebox{0.5\textwidth}{!}{%
    \begin{tabular}{lcccc}
        \toprule
        \diagbox{Statistics}{Datasets} & PISA-Science &PISA-Read  &PISA-Math \\
        \midrule
        \texttt{\#} of exercise records & 7,585,349 & 811,136 &2,594,902  \\
        \texttt{\#} of correct answers &3,691,123 & 528,570 &1,206,973\\
        \texttt{\#} of students & 233,405 & 26,881  & 116,805 \\
        \texttt{\#} of exercises & 184 & 103  & 81  \\
        \texttt{\#} of KCs & 21 & 10  & 11  \\
        Average  of  exercise  &32.50 &30.18  &22.22   \\    
        \bottomrule
    \end{tabular}
    }
    \label{table2}
\end{table}

\section{EXPERIMENTS}



\begin{table*}[htbp]
\centering
\caption{\textbf{Results on student performance prediction across different datasets.}} 
\resizebox{0.8\textwidth}{!}{%
\tiny
\begin{tabular}{c|ccc|ccc|ccc}  
\bottomrule
\multirow{2}{*}{Methods} & \multicolumn{3}{c|}{\centering PISA-Science}   & \multicolumn{3}{c|}{\centering PISA-Read} & \multicolumn{3}{c}{\centering PISA-Math} \\
&  AUC  & ACC & RMSE   &  AUC  & ACC & RMSE   &  AUC  & ACC & RMSE\\
\bottomrule
IRT               & 62.21 & 61.50 & 0.559 & 63.80 & 62.26 & 0.544 & 64.33 & 62.24 & 0.542 \\
\textbf{HCD-IRT}           & \textbf{65.10} & \textbf{63.34} & \textbf{0.536} & \textbf{66.42} & \textbf{65.37} & \textbf{0.529} & \textbf{67.50} & \textbf{66.50} & \textbf{0.524} \\
-~CEA             & 63.50 & 62.10 & 0.548 & 64.55 & 63.46 & 0.532 & 65.47 & 64.37 & 0.528 \\
-~RSA             & 64.80 & 63.08 & 0.543 & 65.24 & 64.28 & 0.531 & 66.38 & 65.26 & 0.525 \\
\hline
MIRT              & 66.54 & 63.54 & 0.524 & 67.76 & 64.73 & 0.519 & 66.84 & 64.20 & 0.516 \\
\textbf{HCD-MIRT}          & \textbf{69.78} & \textbf{67.47} & \textbf{0.507} & \textbf{70.87} & \textbf{68.55} & \textbf{0.504} & \textbf{69.37} & \textbf{68.44} & \textbf{0.502} \\
-~CEA             & 67.59 & 65.29 & 0.512 & 68.27 & 66.84 & 0.514 & 67.75 & 66.57 & 0.513 \\
-~RSA             & 68.12 & 66.76 & 0.510 & 69.58 & 67.46 & 0.512 & 68.19 & 67.68 & 0.505 \\
\hline
DINA              & 67.67 & 64.83 & 0.547 & 68.92 & 65.71 & 0.539 & 68.15 & 65.47 & 0.533 \\
\textbf{HCD-DINA}          & \textbf{72.34} & \textbf{70.29} & \textbf{0.535} & \textbf{73.62} & \textbf{71.68} & \textbf{0.533} & \textbf{72.93} & \textbf{70.84} & \textbf{0.524} \\
-~CEA             & 70.82 & 68.19 & 0.542 & 71.47 & 69.42 & 0.526 & 71.13 & 68.64 & 0.528 \\
-~RSA             & 71.96 & 69.73 & 0.537 & 72.73 & 70.59 & 0.524 & 72.12 & 69.21 & 0.526 \\
\hline
NCDM                   & 78.91 & 71.18 & 0.434 & 79.21 & 74.38 & 0.417 & 80.61 & 72.69 & 0.424 \\
\textbf{HCD-NCDM}      & \textbf{81.16} & \textbf{73.21} & \textbf{0.422} & \textbf{82.17} & \textbf{76.24} & \textbf{0.401} & \textbf{83.66} & \textbf{75.49} & \textbf{0.411} \\
-~CEA           & 78.99 & 71.45 & 0.431 & 79.94 & 74.71 & 0.416 & 81.25 & 73.14 & 0.421 \\
-~RSA           & 79.48 & 72.15 & 0.427 & 81.45 & 75.58 & 0.408 & 81.37 & 73.95 & 0.417 \\
\bottomrule
\end{tabular}
}
\label{tableMain}
\end{table*}

In this section, we will introduce the real dataset used for the experiments and provide detailed descriptions of key aspects of the training process and the selected baseline models. We will then compare the impact of incorporating the HCD features proposed in this study on student performance in relation to the baseline models. To clearly demonstrate the effects of the HCD hierarchy constraints on model performance and knowledge state assessment, we will delve into the following research questions:

\begin{itemize}
\item RQ1: How does the proposed HCD framework impact student performance prediction compared to existing frameworks?
\item RQ2: Can the proposed HCD framework constrain the modeling of group knowledge state?
\item RQ3: How does the proposed HCD framework influence the modeling of individual knowledge state?
\item RQ4: What is the interpretability of the diagnostic results from the HCD framework?
\item RQ5: What is the contribution of hierarchical features and personalized features to students' cognitive understanding?
\item RQ6: How does HCD enhance the diagnostic effectiveness for students?
\end{itemize}

\subsection{Datasets}

Under the Hierarchy Constraint-Aware Cognitive Diagnosis framework, we conducted extensive subject tests to comprehensively validate the effectiveness of the proposed model. The dataset used is the publicly available 2015 Programme for International Student Assessment (PISA 2015) dataset\footnote{https://www.oecd.org/en/about/programmes/pisa/pisa-data.html}, which covers over 500,000 students from 73 countries and regions. We extracted three subsets from this dataset: science, reading, and mathematics. Notably, we classified items marked as "Full credit" in the dataset as correct responses (labelled as 1), while "No credit" and "Partial credit" were considered incorrect responses (labelled as 0). All other cases were treated as unanswered questions. 

To ensure the reliability of the analysis results, we excluded student records with fewer than 30 entries, and for the mathematics dataset, we excluded records with fewer than 20 entries. Table \ref{table2} presents the basic statistical information for these datasets. Finally, we randomly divided all datasets into 70\% for training, 10\% for validation, and 20\% for testing to ensure the effectiveness of model evaluation.

\subsection{Experimental Setup}

At the beginning of the experiments, we initialized the parameters using Xavier initialization and employed a 5-fold cross-validation method for training, with the final results representing the average of the 5 folds. The experimental environment consisted of a 64-bit Ubuntu 20.04.5 LTS server equipped with a 2.30GHz Intel Xeon Gold 5218 CPU and a 32GB Tesla V100 GPU, with the computations based on the PyTorch framework.

\subsection{Baselines}


    



    
To assess the impact of hierarchy constraint perception on model performance, we proposed four implementation schemes based on the HCD framework, incorporating typical cognitive diagnosis methods: HCD-IRT, HCD-MIRT, HCD-DINA, and HCD-NCDM. During the implementation process, we replicated these models \footnote{https://github.com/bigdata-ustc/EduCDM} and conducted fine-tuning of parameters to ensure that each model operates at its optimal state, thus ensuring the fairness of the comparative results. We used area under the curve (AUC), accuracy (ACC), and root mean square error (RMSE) as the main performance metrics to comprehensively evaluate the effectiveness of each model.

\begin{figure*}
    \centering
    \includegraphics[width=1.0\textwidth]{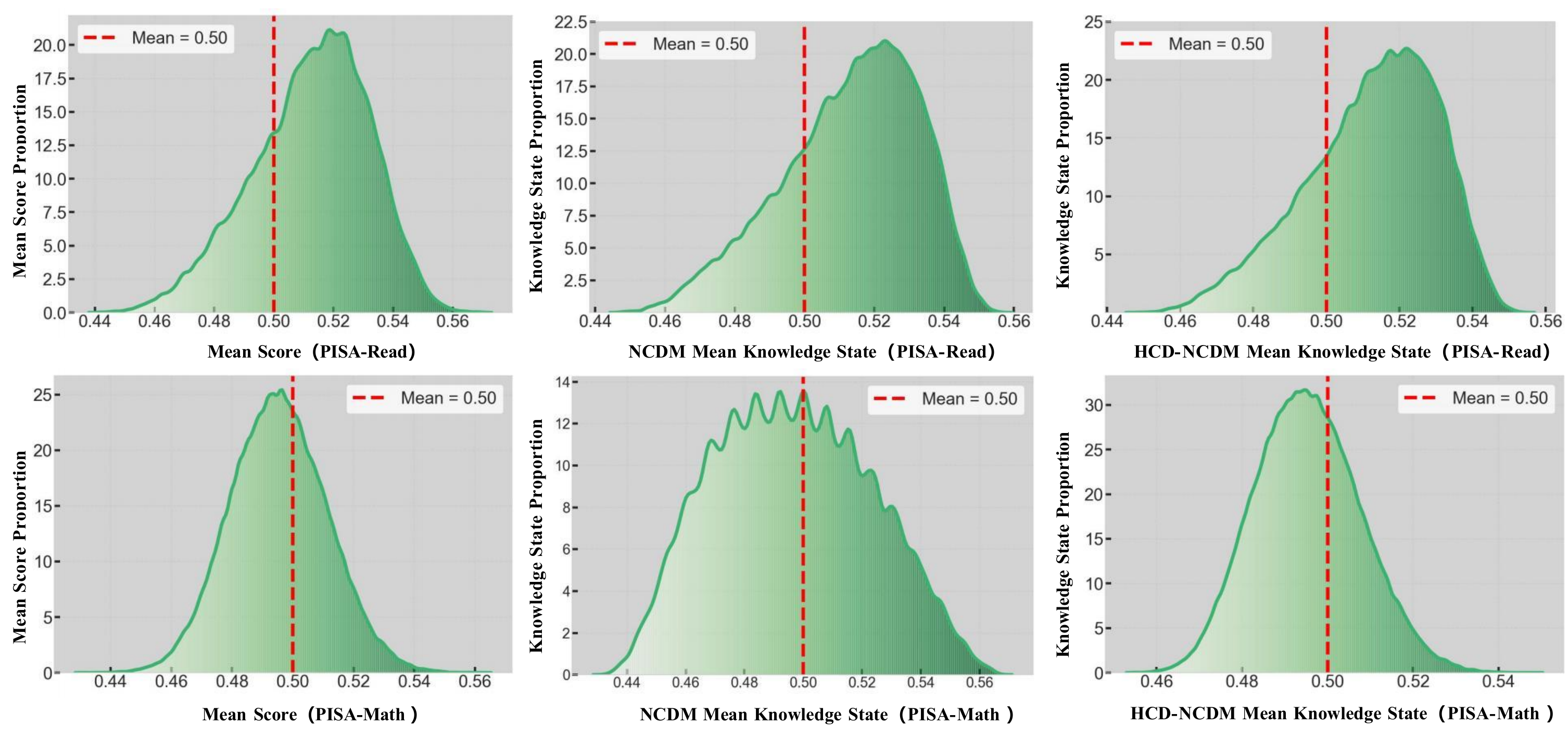}
    \caption{\textbf{Comparison of kernel density plots for the mean score based on prior statistics, the mean knowledge state of students from the NCDM model, and the mean knowledge state of students from the HCD-NCDM model on the PISA-Read and PISA-Math datasets.}}\label{figureMath}
\end{figure*}

\begin{itemize}

\item \textbf{IRT} \citep{bib020}: This statistical model is used to assess students' mastery of specific knowledge concepts. By analyzing students' performance on different test items, the IRT model can estimate students' ability levels and the difficulty of the items.

\item \textbf{MIRT} \citep{bib021}: As an extension of Item Response Theory (IRT), the MIRT model is used to evaluate students' performance across multiple latent ability dimensions. By analyzing student responses to different items, the MIRT model can estimate multiple ability dimensions and item parameters simultaneously.

\item \textbf{DINA} \citep{bib038}: The DINA model assumes that each item involves multiple knowledge concepts and infers students' mastery of these points by analyzing their response patterns. The model uses binary classification (mastered or not mastered) to describe the mastery status of each knowledge concept.

\item \textbf{NCDM} \citep{bib047}: The NCDM model utilizes neural networks to learn the complex relationships between students and items. Through multi-layer neural networks, this model achieves accurate and interpretable cognitive diagnosis, replacing traditional manually designed predictive functions.

\end{itemize}

\subsection{Performance Prediction (RQ1)}




Due to the inability to directly measure students' true knowledge state, evaluating the performance of cognitive diagnosis models is challenging. Although our primary goal is not to directly enhance the predictive performance of the models, indirectly assessing their validity through student performance predictions can still reveal the significance of hierarchy constraint perception (HCD) in cognitive diagnosis.

Table \ref{tableMain} presents the performance of various classical cognitive diagnosis models combined with HCD. Through analysis, we derive several key conclusions. First, overall, methods that incorporate HCD (such as HCD-IRT) significantly outperform traditional methods (such as IRT), indicating the importance of integrating hierarchy constraint features into cognitive diagnosis. Second, after applying HCD to the NCDM model, we achieved optimal performance in AUC, ACC, and RMSE metrics, further demonstrating that the powerful fitting ability of neural networks can more effectively assess students' knowledge state. Lastly, analysis of the three datasets revealed that HCD exhibits strong adaptability. Whether in the larger PISA-Science dataset or the less interactive PISA-Math dataset, HCD performed well. Even in the relatively smaller PISA-Read dataset, the positive effects of hierarchy constraints were evident, proving the wide applicability of HCD.

To validate the effectiveness of two key modules in the HCD framework (CEA and RSA), we conducted ablation experiments. Specifically, we replaced each layer with a simple aggregation layer that merely averages the inputs while keeping the other layers unchanged. Table \ref{tableMain} lists the results under different scenarios, leading to the following conclusions: First, regardless of which layer was replaced, the final performance decreased, indicating that each module contributes to the overall performance and demonstrating their effectiveness in modeling both intra- and inter-layer relationships. Second, when replacing the convolution-enhanced attention layer within the hierarchy, the performance decline was most significant, highlighting the critical importance of accurately modeling the relationships between the same knowledge concepts within the same hierarchy in hierarchy constraint modeling.

\subsection{Hierarchy Constraint Aware (RQ2)}




The introduction of hierarchy constraint perception aims to enable existing cognitive diagnosis models to more accurately reflect students' relative ability performance, thus allowing for a more reasonable inference of their ability levels across various knowledge concepts. This constraint helps ensure that assessment results remain within a reasonable range. Figure \ref{figureMath} illustrates the performance effects of hierarchy constraints on two typical datasets, PISA-Read and PISA-Math, using kernel density plots.

Several important findings emerge. First, overall, both the NCDM and the hierarchy constraint-enhanced HCD-NCDM models are able to effectively distinguish students' ability distributions on both datasets, aligning well with the average trends of students' prior distributions. This indicates the effectiveness of current cognitive diagnosis techniques in assessing students' knowledge state. Second, a comparison of the two datasets reveals that, on the smaller PISA-Read dataset, while the average knowledge state levels learned by both models are similar to the students' prior distribution, the overall distribution significantly deviates from a normal distribution. In contrast, on the larger PISA-Math dataset, the distribution of students' abilities is closer to a normal distribution, which aligns with the general of ability development in human educational contexts\citep{bib062}.

Third, from the experimental results, the NCDM model on the PISA-Math dataset shows considerable fluctuations in the learned knowledge state of students. In contrast, after incorporating hierarchy constraint perception, students' ability performances become smoother and exhibit a higher degree of fit with the prior statistical distribution. This indicates that the cognitive diagnosis model, after the introduction of hierarchy constraints, not only learns more reasonable student abilities but also avoids neglecting relative ability performance solely to increase model accuracy, resulting in better interpretability. Finally, further observation of the two datasets, particularly regarding the distribution of students with an ability level of 0.50, reveals that the model with hierarchy constraint perception aligns more closely with the prior distribution. For example, in the PISA-Math dataset, students with an ability level of 0.50 constitute about 24.0\% of the prior distribution, while the NCDM model only learns 13.5\%; conversely, the HCD-NCDM model accounts for 28.0\% of students at this level, which is closer to the prior distribution. This again demonstrates the positive impact of incorporating hierarchy constraint perception in enabling cognitive diagnosis models to more accurately reflect students' knowledge state.

\subsection{Individual Knowledge State Modeling (RQ3)}
\begin{figure}
    \centering
    \includegraphics[width=0.5\textwidth]{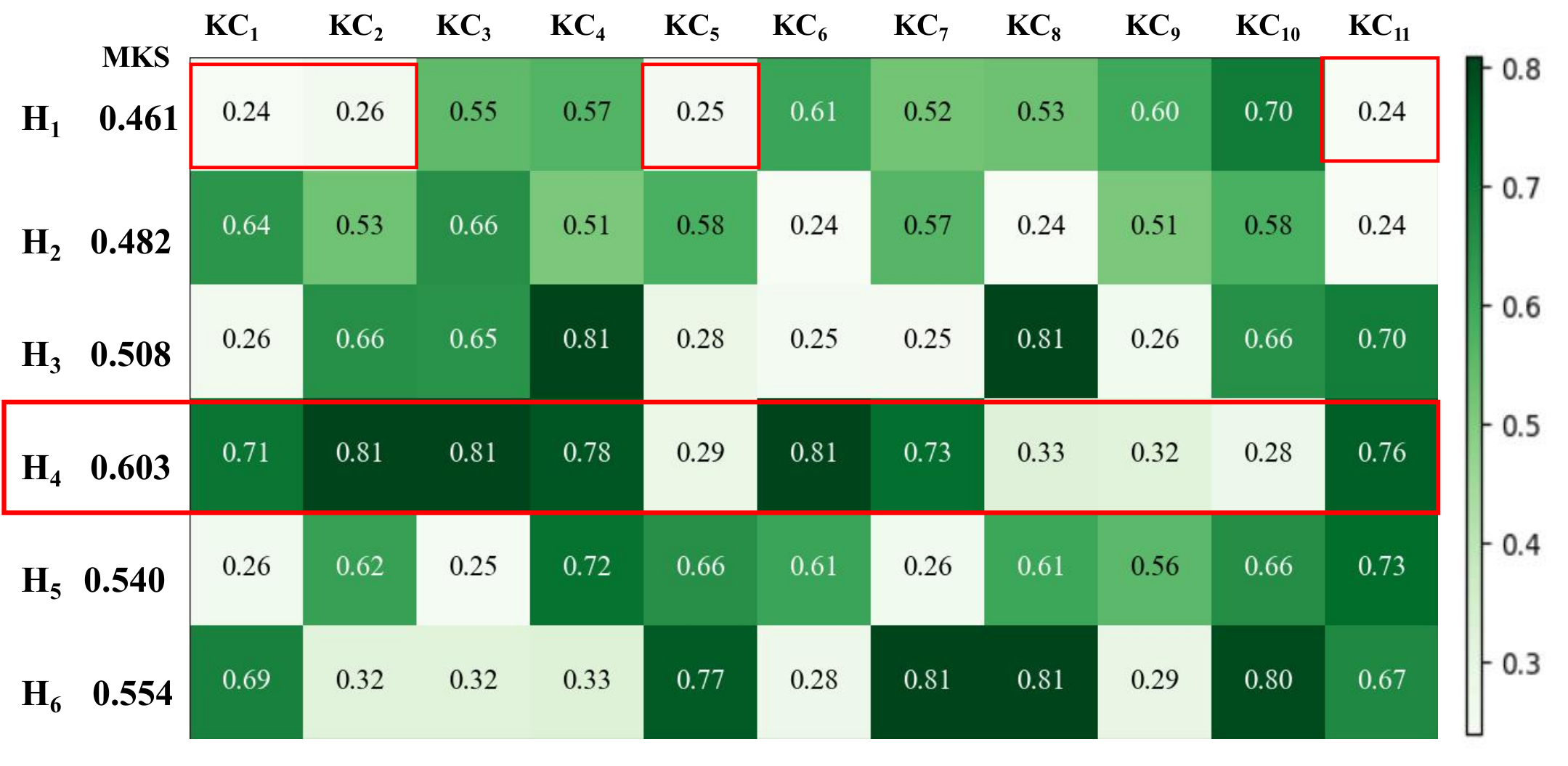}
    \caption{\textbf{Demonstrating knowledge concept Mastery for Students Across Six Levels in the PISA-Math Dataset: Random Selection of One Student from Each Level}\label{figureheat}}
\end{figure}

\begin{figure}
    \centering
    \includegraphics[width=0.50\textwidth]{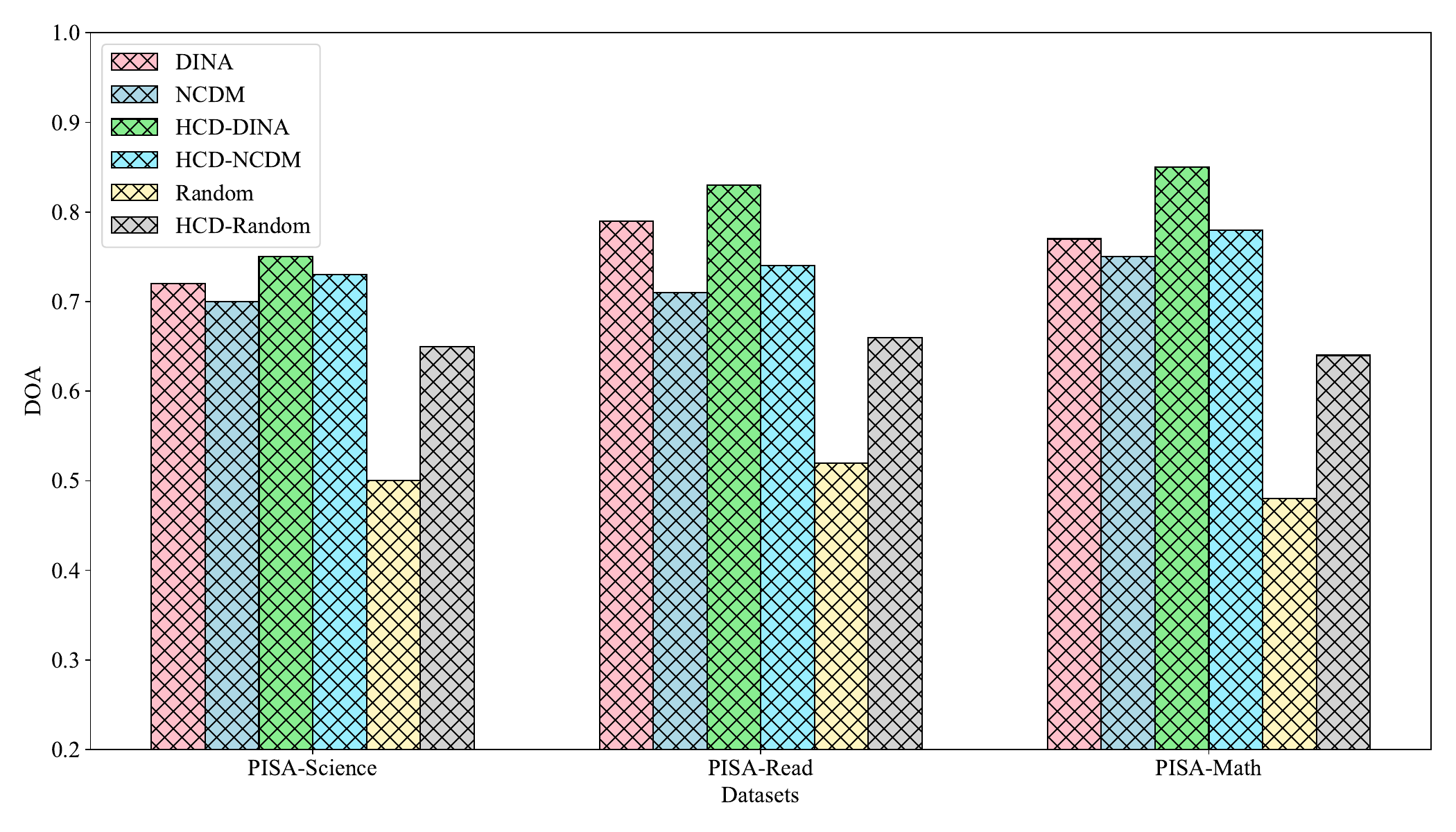}
    \caption{\textbf{Comparative analysis of model DOA interpretability across three distinct datasets.}\label{figureDOA}}
\end{figure}



Assessing the accuracy of student knowledge state is crucial for cognitive diagnosis. Figure \ref{figureMath} illustrates the hierarchy constraints under group effects, but the extent to which individuals are influenced by these constraints requires further exploration. We categorized students into six levels \citep{bib008} and randomly selected one student from each level to visually present their mastery of 11 KCs in the PISA-Math dataset (see Figure \ref{figureheat}), which reveals several key observations.

First, there are significant differences in knowledge state among students at different hierarchies. Each student's answering records vary, resulting in diverse levels of knowledge mastery. This indicates that the model has strong discriminative power in capturing students' knowledge state. Second, regardless of the hierarchy, certain knowledge concepts consistently show low mastery. For instance, in the H1 level, several knowledge concepts have mastery levels close to 0.24. This may be due to students not having encountered these knowledge concepts; nevertheless, the powerful fitting capability of the neural network still assigns some level of mastery, demonstrating its explanatory power in cognitive diagnosis, though there may also be some bias. Finally, the analysis found that individual knowledge mastery does not always align with expectations. For example, the average mastery level of students at the H4 level did not accurately reflect their expected position between H3 and H5. This suggests that even with the introduction of hierarchy constraints, not all students' knowledge state can be precisely modeled. The complexity of students' knowledge state is influenced by various factors, such as affect during answering \citep{bib048}, incorrect answers, and guessing. Therefore, future work still needs to further optimize cognitive diagnosis models to enhance their capability for accurate modeling of individual knowledge state.

\subsection{Interpretability of Diagnostic Results (RQ4)}





In cognitive diagnosis, a student's mastery of specific knowledge concepts significantly impacts their performance. Specifically, if student  S1  has a better mastery of knowledge concept  k  than student S2 , then S1  is likely to have a significantly higher probability of success when solving problems related to that knowledge concept compared to  S2  \citep{bib049}. Based on this logic, we introduce the Degree of Agreement (DOA) \citep{bib050} as a metric to evaluate the effectiveness of different models in explaining student performance.

\begin{figure*}
    \centering
    \includegraphics[width=0.85\textwidth]{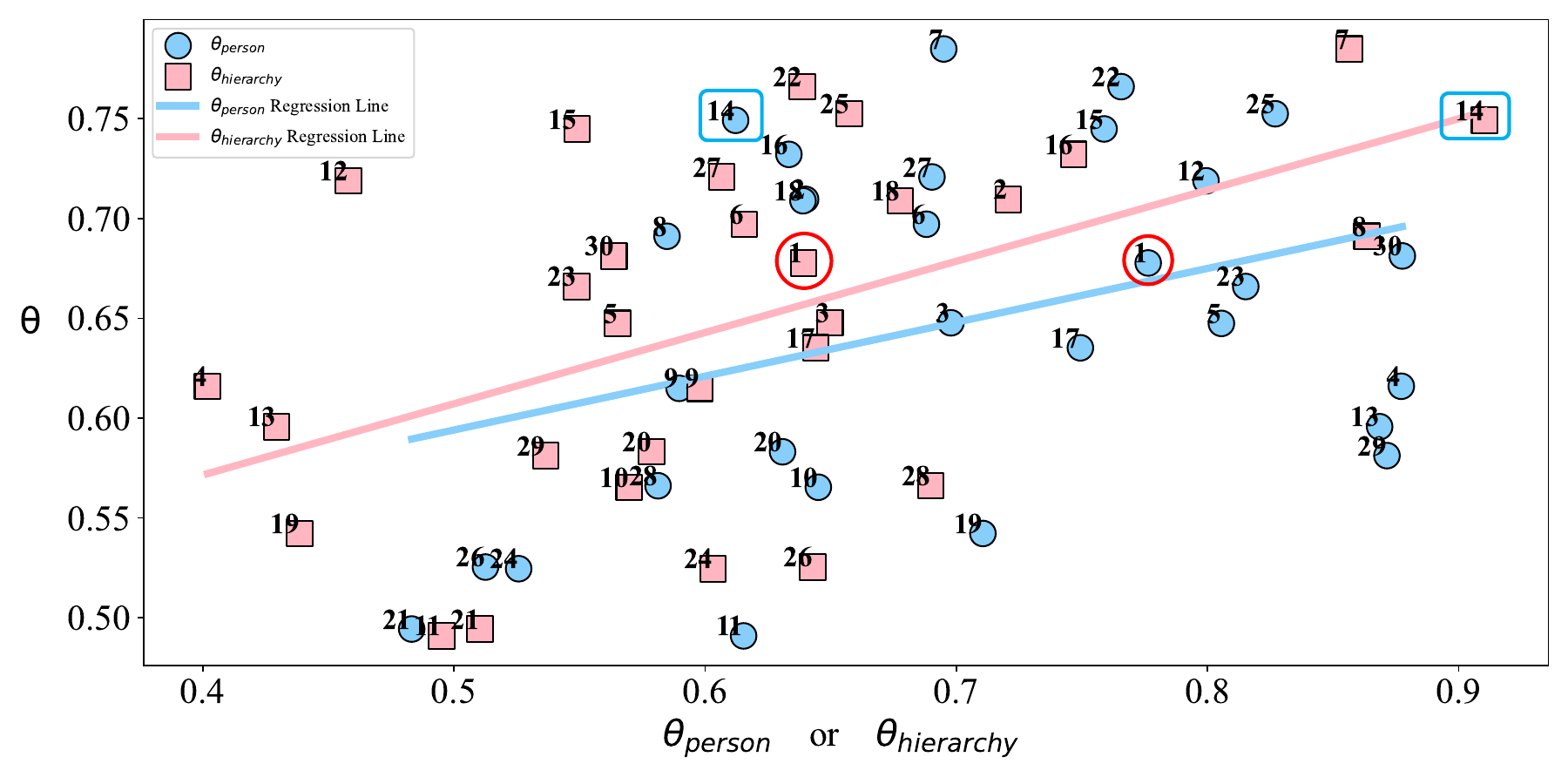}
    \caption{\textbf{Randomly selecting 30 students from the PISA-Science dataset and averaging their abilities across various knowledge concepts, we perform the fusion of the two ability characteristics according to Equation Eq.(17).}\label{figureθ}}
\end{figure*}

Traditional benchmark models, such as Item Response Theory (IRT) and Multidimensional Item Response Theory (MIRT), typically rely on latent trait vectors to depict student abilities; however, this approach often fails to reveal the specific mastery levels of students on particular knowledge concepts. To overcome this limitation, this study compares the hierarchy Constraint Awareness model (HCD) with models like Random, DINA, and NCDM, focusing on their performance on the DOA metric. As shown in Figure \ref{figureDOA}, the DOA results for each model provide important insights for evaluating their interpretability. Through this analysis, we aim to gain a clearer understanding of the differences among models in describing students' knowledge mastery levels.

From the data presented in the figure, we can draw several key conclusions. First, the DINA model, without hierarchy constraint awareness, exhibits the highest DOA values across all three datasets, demonstrating its superior ability to explain student knowledge state. This result is closely linked to the theoretical framework on which the DINA model is based, highlighting its clear advantage in interpretability. However, despite its strong theoretical performance, the DINA model still falls short in practical applications regarding the predictive effectiveness of student responses, failing to provide comprehensive support.

Second, all classical cognitive diagnostic models incorporating hierarchy constraint awareness outperform their traditional counterparts, further emphasizing the effectiveness of our proposed approach in enhancing model interpretability. Lastly, an intriguing observation is that the Random model, based on random assignments, shows relatively stable performance on the DOA metric, maintaining around 0.50, which aligns with our expectations. However, the HCD-Random model, with the introduction of hierarchy awareness, shows a significant improvement in its DOA value. This indicates that even relying solely on hierarchy constraint awareness for assessing student knowledge state can yield satisfactory results, thereby underscoring the important influence of hierarchy relative ability learning on the development of students' knowledge state.

\subsection{Visualized Diagnosis Enhancement (RQ5)}

The interaction between hierarchical abilities features and personalized ability features, as well as which feature has a greater impact on students, is visually illustrated in Figure \ref{figureθ}. First, we observe that each student's personalized ability \(\theta_{person}\) and hierarchical abilities \(\theta_{hierarchy}\) show significant differences, highlighting the importance of personalized student ability modeling. Second, the influence weights of personalized and hierarchical abilities on students vary. For example, for Student S1, the personalized ability level is higher than the hierarchical abilities level, yet hierarchical abilities plays a greater role in the final student ability \(\theta\). Conversely, for Student S14, although the hierarchical abilities level is higher, the influence weight of personalized ability on overall ability is greater. This indicates that each student's ability is affected differently by the two features. Finally, we note that the slope of the regression line for hierarchical abilities is steeper than that for personalized ability, suggesting that in the current performance of the 30 students, hierarchical abilities has a greater overall impact on final student ability than personalized ability.

\subsection{Case Study (RQ6)}

\begin{figure}
    \centering
    \includegraphics[width=0.50\textwidth]{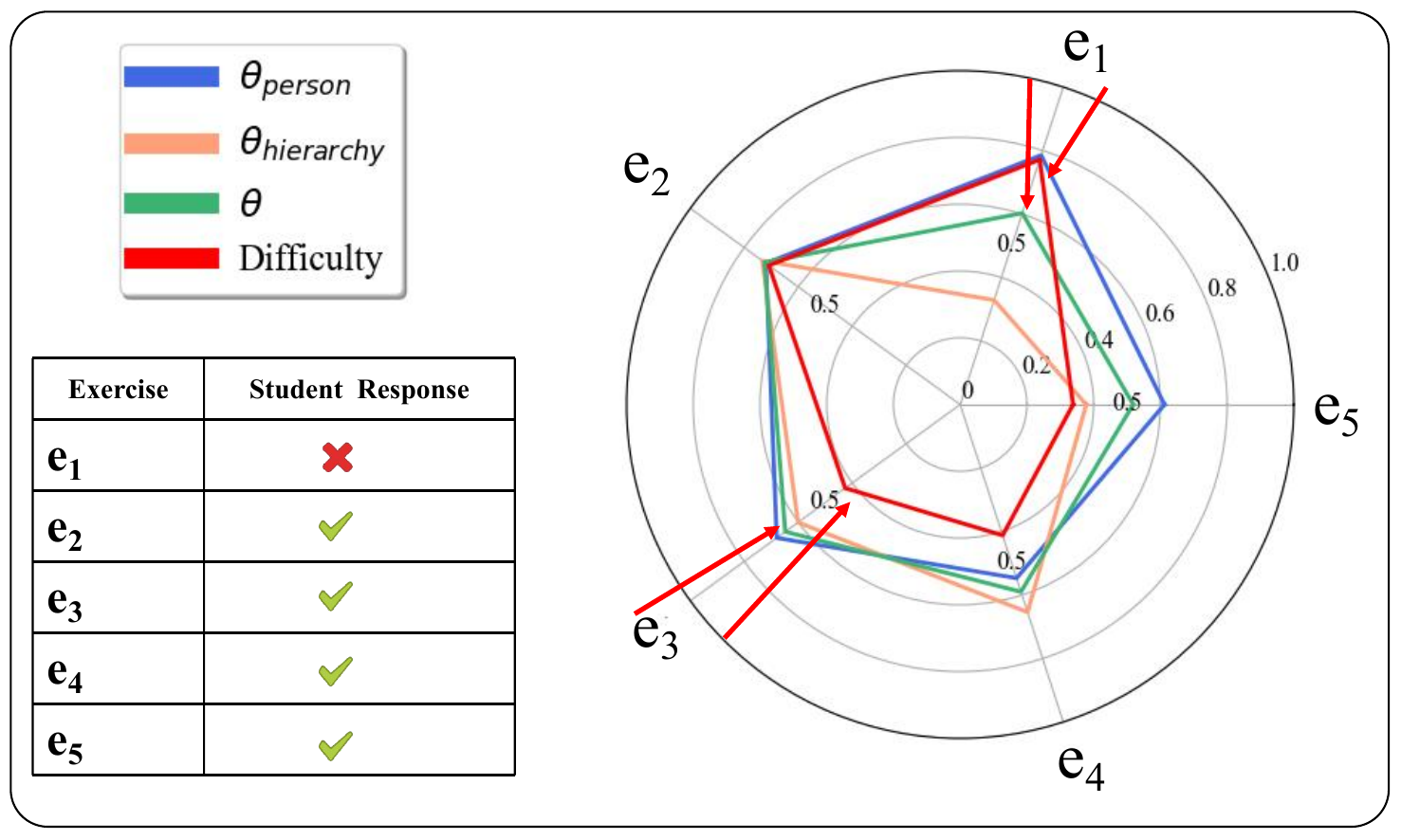}
    \caption{\textbf{HCD-NCDM cognitive diagnosis case study: interrelationships among personalized ability \(\theta_{person}\), hierarchical abilities \(\theta_{hierarchy}\), Overall Ability \(\theta\), and Item Difficulty.}\label{figurecase}}
\end{figure}


The relationship between a student's cognitive diagnosis results, their abilities, and item difficulty can be quantitatively analyzed using Eq.(19). We randomly selected one student from the PISA-Science dataset along with five items they answered. The left half of Figure \ref{figurecase} displays the student's response results, while the right half presents the model's diagnostic outcomes. The radar chart illustrates the student's mastery of various knowledge concepts alongside the difficulty of the items.

From the figure, it is evident that when a student's knowledge mastery meets the requirements of the item, they are more likely to provide a correct answer. For instance, in item e3, the student's personalized ability \(\theta_{person}\) and hierarchical ability \(\theta_{hierarchy}\) are 0.64 and 0.60, respectively, while the overall ability \(\theta = 0.63\) and the item's difficulty is 0.41. Since the student's proficiency in the knowledge concepts related to this item exceeds the required level, they successfully answered it. Conversely, in item e1, the student's personalized ability \(\theta_{person}\) is 0.795, and their hierarchical ability \(\theta_{hierarchy}\) is 0.37, with an overall ability of \(\theta = 0.60\) and the item's difficulty at 0.793. Although the student's personalized ability is higher than the item difficulty, their overall ability falls short of what is needed to solve the item, resulting in an incorrect response. This phenomenon aligns with the student's answers and further illustrates how incorporating hierarchical abilities can significantly enhance the accuracy of predictions regarding student responses.

\section{Conclusions and Limitations}

In this paper, we propose a novel Hierarchy Constraint-Aware cognitive Diagnosis (HCD) framework aimed at quantitatively analyzing students' cognitive state. Specifically, we designed a two-stage solution. In the hierarchy constraint awareness modeling phase, we introduced a convolution-enhanced attention network for the same knowledge concepts at the same level, while a feature-sampling attention network was used to simulate the impact of hierarchy constraint awareness on knowledge mastery between levels. In the personalized diagnosis enhancement phase, we combined personalized knowledge proficiency (absoluteness) with the knowledge proficiency influenced by hierarchy constraint awareness (relativeness).

Within this framework, we implemented four specific models (i.e., HCD-IRT, HCD-MIRT, HCD-DINA, and HCD-NCDM) and conducted extensive experiments on real-world datasets to validate the effectiveness and interpretability of the HCD framework. Finally, we analyzed and discussed the impact of hierarchy constraints on both student groups and individuals, with the hope that this research will inspire further exploration in related areas.

However, we also recognize that the current hierarchy constraint awareness primarily targets effective modeling of students' overall levels, and finer-grained hierarchy modeling based on knowledge proficiency still requires further investigation. Additionally, the learning process of students is a complex educational psychology issue; relying solely on students' ability characteristics and item features makes it challenging to comprehensively reveal the relationship between their knowledge state and cognitive levels.



\bibliographystyle{plainnat}
\bibliography{HCD.bib}

\end{document}